\documentclass[acmsmall]{acmart}

\usepackage{graphicx}
\usepackage[utf8]{inputenc}
\usepackage{tabularx}
\usepackage{multirow}
\usepackage{booktabs}
\usepackage{rotating}

\AtBeginDocument{%
  \providecommand\BibTeX{{%
    \normalfont B\kern-0.5em{\scshape i\kern-0.25em b}\kern-0.8em\TeX}}}

\setcopyright{acmlicensed}
\acmPrice{15.00}
\acmDOI{10.1145/3611054}
\acmYear{2023}
\copyrightyear{2023}
\acmSubmissionID{play23main-p7374-p}
\acmJournal{PACMHCI}
\acmVolume{7}
\acmNumber{CHI PLAY}
\acmArticle{408}
\acmMonth{11}
\received{2023-02-21}
\received[accepted]{2023-07-07}

\begin{document}

\title[Auto-Paizo Games: Towards Understanding the Design of Games That Aim to Unify a Player’s Physical Body (...)]{Auto-Paizo Games: Towards Understanding the Design of Games That Aim to Unify a Player’s Physical Body and the Virtual World}

\author{Rakesh Patibanda}
\orcid{0000-0002-2501-9969}
\email{rakesh.patibanda@monash.edu}
\affiliation{%
  \institution{Monash University}
  \department {Exertion Games Lab, Department of Human-Centred Computing}
  \city{Melbourne}
  \country{Australia}
}

\author{Chris Hill}
\orcid{0000-0003-2625-0254}
\email{chrishillcs@gmail.com}
\affiliation{%
  \institution{Monash University}
  \department {Exertion Games Lab, Department of Human-Centred Computing}
  \city{Melbourne}
  \country{Australia}
}

\author{Aryan Saini}
\orcid{0000-0002-2844-3343}
\email{aryan@exertiongameslab.org}
\affiliation{%
  \institution{Monash University}
  \department {Exertion Games Lab, Department of Human-Centred Computing}
  \city{Melbourne}
  \country{Australia}
}

\author{Xiang Li}
\orcid{0000-0001-5529-071X}
\email{xl529@cam.ac.uk}
\affiliation{%
  \institution{University of Cambridge}
  \department{Department of Engineering}
  \city{Cambridge}
  \country{UK}
}

\author{Yuzheng Chen}
\orcid{0000-0001-9538-5369}
\email{yuzheng.chen18@student.xjtlu.edu.cn}
\affiliation{%
  \institution{Xi’an Jiaotong-Liverpool University}
  \city{Suzhou}
  \country{China}
}

\author{Andrii Matviienko}
\orcid{0000-0002-6571-0623}
\email{matviienko.andrii@gmail.com}
\affiliation{%
  \institution{KTH Royal Institute of Technology}
  \city{Stockholm}
  \country{Sweden}
}
\affiliation{%
  \institution{Monash University}
  \department {Exertion Games Lab, Department of Human-Centred Computing}
  \city{Melbourne}
  \country{Australia}
}

\author{Jarrod Knibbe}
\orcid{0000-0002-8844-8576}
\email{jarrod.knibbe@unimelb.edu.au}
\affiliation{%
  \institution{University of Melbourne}
  \city{Melbourne}
  \country{Australia}
}

\author{Elise van den Hoven}
\orcid{0000-0002-0888-1426}
\email{elise.vandenhoven@uts.edu.au}
\affiliation{%
  \institution{University of Technology Sydney}
  \city{Sydney}
  \country{Australia}
}
\affiliation{%
  \institution{Eindhoven University of Technology}
  \city{Eindhoven}
  \country{Netherlands}
  }

\author{Florian ‘Floyd’ Mueller}
\orcid{0000-0001-6472-3476}
\email{floyd@exertiongameslab.org}
\affiliation{%
  \institution{Monash University}
  \department {Exertion Games Lab, Department of Human-Centred Computing}
  \city{Melbourne}
  \country{Australia}
}

\renewcommand{\shortauthors}{Patibanda, Hill, Saini, Li, Chen, Matviienko, Knibbe, van den Hoven and Mueller.}

\begin{abstract}
  Most digital bodily games focus on the body as they use movement as input. However, they also draw the player’s focus away from the body as the output occurs on visual displays, creating a divide between the physical body and the virtual world. We propose a novel approach – the "Body as a Play Material" – where a player uses their body as both input and output to unify the physical body and the virtual world. To showcase this approach, we designed three games where a player uses one of their hands (input) to play against the other hand (output) by loaning control over its movements to an Electrical Muscle Stimulation (EMS) system. We conducted a thematic analysis on the data obtained from a field study with 12 participants to articulate four player experience themes. We discuss our results about how participants appreciated the engagement with the variety of bodily movements for play and the ambiguity of using their body as a play material. Ultimately, our work aims to unify the physical body and the virtual world.
\end{abstract}




\maketitle

\section{Introduction}\label{sec:intro}
As a community, we are increasingly interested in creating digital bodily games using novel movement-sensing technologies, such as Nintendo Wii and Ring Fit (e.g., \cite{Benford_Ramchurn_Marshall_2020,Mueller_Gibbs_Vetere_Edge_2017,Mueller_Isbister_2014}). These games promote physical activity \cite{Bianchi-Berthouze_2013,Isbister_Mueller_2015}, help develop bodily awareness \cite{Marquez_Segura_2013} and foster knowledge of the inner self \cite{Pasch_Bianchi_2009,de_Vignemont_2020}. However, while these games utilise the body as input, visual displays typically provide output, creating a fundamental division between the physical body and the virtual world \cite{Mueller_Isbister_2014}. As a result, it has been suggested that players continually shift their focus, negatively impacting their overall bodily experience \cite{Marquez_Segura_2013}.


\begin{figure}[h]
  \centering
  \includegraphics[width=0.95\linewidth]{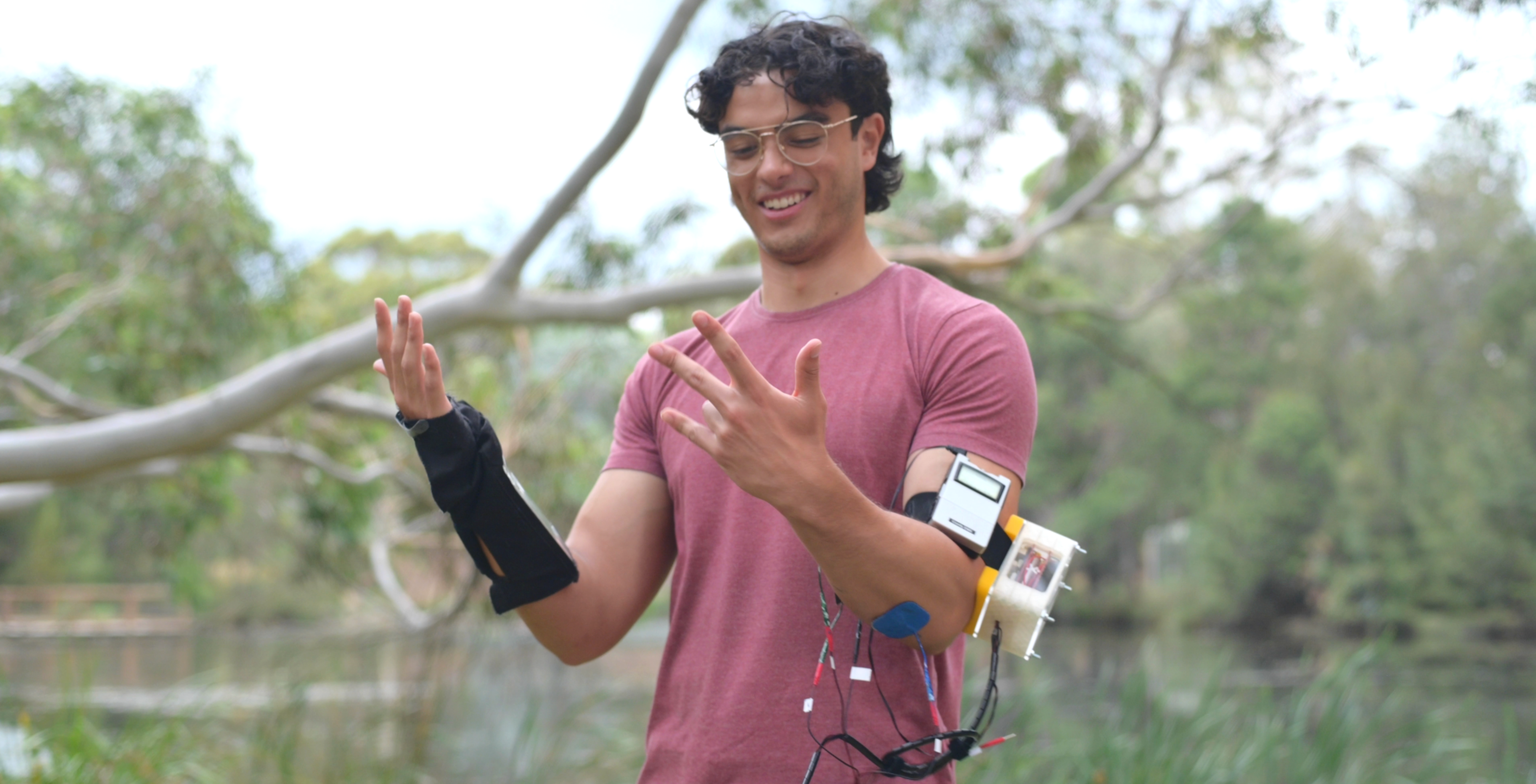}
  \caption{A participant is playing “Elements”. The non-EMS hand (the player’s right hand) shows the “Water” symbol, while the EMS hand (the player’s left hand) shows the “Air” symbol caused by actuation and therefore closing the ring finger. The rules state that “Air” beats “Water”; hence, the EMS hand wins.}
  \Description{A participant is playing “Elements”. The non-EMS hand (the player’s right hand) shows the “Water” symbol, while the EMS hand (the player’s left hand) shows the “Air” symbol caused by actuation and therefore closing the ring finger. The rules state that “Air” beats “Water”; hence the EMS hand wins.}
  \label{fig:teaser}
\end{figure}

In response, game design researchers have called out the need for deeper integration of the human body and computers to experience “the body as play” \cite{Mueller_Byrne_Andres_Patibanda_2018,Mueller_Kari_Li_Wang_Mehta_Andres_Marquez_Patibanda_2020}. The community responded by designing games that, for example, use the body’s surface as an interactive screen \cite{Chatain_Sisserman_Reichardt_Fayolle_Kapur_Sumner_Zünd_Bermano_2020} and allow users to play with their interior body, such as their body temperature \cite{Li_Patibanda_Brandmueller_Wang_Berean_Greuter_Mueller_2018}. While these games begin closing the divide between the body and the virtual world, the player still focuses on information external to their body, “using” their body to play “with” a screen. In response, we asked ourselves: \textit{How, then, can we design games that allow a player to focus solely on themselves, where they “use” their body to play “with” the body?}

We propose a novel approach – the “Body as a Play Material” – where a player uses their body to play with their body by using it as input and output by loaning bodily control to a computer~\cite{Floyd_Mueller_Patibanda_Byrne_Li_Wang_Andres_2021}. To showcase our approach, we designed three novel bodily games called "Elements", "Numbers", and "Slap-Me-If-You-Can", inspired by traditional hand games like rock-paper-scissors. We call this suite of games “Auto-Paízo” (Self-Play in Greek) games. Traditional hand games usually require two-players; however, thanks to our approach, such games can also be played by a single player. In our games, a player plays with one of their hands (as input) against their other Electrical Muscle Stimulation (EMS)-controlled hand (as output) (Fig. \ref{fig:teaser}). These games not only showcase a novel way to experience bodily play, but could also be employed in the medical domain, for example, for motor rehabilitation programs, where patients can engage with interactive games that facilitate the recovery of motor skills \cite{Lohse_Hilderman_2014,Panaite_Rişteiu_Olar_Leba_Ionica_2019}.

We obtained ethics from our institution to conduct a seven-day field study \cite{Grzyb_Dolinski_2021} of our games and interviewed the 12 participants. We conducted a field study to allow our participants to experience the games outside the lab’s environment (see participants’ engagement patterns in section \ref{sec5:quantitativeresults}), while also adhering to COVID-19 restrictions. By analysing the interview data using inductive thematic analysis, we articulated four player experience themes: 1) loaning bodily control to a computer, 2) playfully engaging with the integrated computer, 3) playing with the ambiguity of the computer-controlled body parts, and 4) bodily awareness when loaning bodily control to a computer to play. We found that our participants enjoyed engaging with the variety of bodily movements for play and the ambiguity of using their body as a play material. This supports prior theories that varied bodily movements \cite{Marquez_Segura_2013} and ambiguity \cite{Gaver_Beaver_Benford_2003,Sutton-Smith_2001} are key concepts in the design of bodily games \cite{Marquez_Segura_2013,Mueller_Isbister_2014}. We extend these theories by reflecting on our themes to articulate design considerations for designing bodily games by using our approach. We make three contributions: 

\begin{itemize}
  \item A system contribution by presenting three games, adding to HCI’s collection of novel systems \cite{Wobbrock_Kientz_2016}. This contribution can inspire industry developers to utilise EMS through the lens of entertainment \cite{Lopes_Baudisch_2017}, possibly to benefit the medical domain by creating engaging EMS experiences.
  \item An empirical contribution by presenting results in the form of four descriptive user experience themes from our study. This contribution can be useful for user experience researchers who aim to understand the potential of using the Body as a Play Material approach.
  \item A practical contribution by articulating design considerations by reflecting on our results to create future games using our approach. This contribution can be helpful for game design researchers interested in creating bodily play experiences that aim to unify the physical body and the virtual world.
\end{itemize}

\section{Related Work}\label{sec2:RW}
Our work was inspired and informed by how the body is the focus of play in many non-digital bodily games, how the game design community is amplifying the human body’s involvement in digital bodily games, and how the body can be used as input and output with the help of Electrical Muscle Stimulation (EMS).

\subsection{Body as the Focus of Gameplay in Non-digital Bodily Games}\label{sec2.1:bodyasfocus}

In many traditional non-digital bodily games, such as rock-paper-scissors, the body is the focus of play \cite{mueller2023towards}. Here, the body is the primary “play material,” and game rules are crafted such that players must employ their bodies interactively within the game and with other players \cite{Caillois_2001,Huizinga_2016}. This pivotal role of the body gives rise to a wide variety of bodily movements, from fine gestures (showing “scissors”) that require precise motor control and quick reactions to the opponent’s movements \cite{Walker_Walker_2004}, to more expansive gross motor movements, such as when playing Hopscotch \cite{Cortazar_2020}. In other words, the player’s body is central to play in these games \cite{Cortazar_2020,Walker_Walker_2004}. We note that players benefit from having complete control over their body when playing such games, adding an element of agency to the gameplay. However, there are also games where players loan bodily control to others. For example, in the “Statues,” one player becomes the “statue” and must hold a pose while other players try to make them laugh or move in other ways \cite{December_2008}. In this game, the body of the “statue” is the focus, while the “statue” player subscribes to the idea that they, at some point, will end up giving control over their body to the other players. In “Statues”, players do not have physical control over each other; however, in a three-legged race, two players physically tie one of their legs together and run together, sharing their movements \cite{Serrano_Zapater_1998}. Similarly, in the game “Trust Fall,” one player falls backwards and is caught by a group of other players, who must work together to ensure the person’s safety \cite{Depping_Mandryk_Johanson_Bowey_Thomson_2016}. These examples demonstrate that while the player’s body is still the focus of gameplay, the control of it can be loaned to others, creating unique game experiences.

In summary, we learned that when simple rules engage with the player’s physical coordination abilities and reaction time, they can facilitate engaging experiences. Therefore, we also focused on simple rules that depend on the player’s physical coordination abilities and reaction time in our games.

\subsection{Amplifying the Human Body’s Involvement in Digital Bodily Games}\label{sec2.1:amplifyingbody}

When it comes to digital games, the role of the human body has been less central, with many games relying on external input devices, such as a keyboard, mouse, or gamepad, to control virtual bodies, like avatars \cite{Mueller_Isbister_2014,Mueller_Kari_Li_Wang_Mehta_Andres_Marquez_Patibanda_2020,Mueller_Matjeka_Wang_Andres_Li_Marquez_Jarvis_Pijnappel_Patibanda_Khot_2020}. However, the emergence of movement-sensing technologies, such as those promoted with the Nintendo Wii and Ring Fit, has enabled the game design community to integrate the body as the primary controller in digital games (e.g., \cite{Mehta_Khot_Patibanda_Mueller_2018,Li_Patibanda_Brandmueller_Wang_Berean_Greuter_Mueller_2018,Montoya_Patibanda_Clashing_Pell_Mueller_2022,Li_VR_Tang_Tong_Patibanda_Mueller_Liang_2021,Li_Captcha_Chen_Patibanda_Mueller_2021,La_Delfa_Baytas_Patibanda_2020}). This amplified the role of the body in play (as input), while providing visual feedback on screens (as output) \cite{Marquez_Segura_2013}. For instance, in the game "Bounden", two players hold the same mobile device: they use their synced movements to guide a ball through a series of levels, resulting in a dance-like experience \cite{oven2014bounden}. The game "Johann Sebastian Joust" uses PlayStation Move controllers to turn the players' bodies into the game controller, with the controllers vibrating and lighting up when the player moves too quickly \cite{McElroy_2012}. We learn from these games that the player's movements and speed can be the primary input for bodily games, which we utilised in our games. However, in both cases, the player's focus remains directed outwards, towards an external display. To foster experiences where the player’s body is the focus of play (which we are interested in), researchers have been exploring ways to incorporate body practices into the experience \cite{Nacke_Kalyn_Lough_Mandryk_2011}. For example, Life Tree \cite{Patibanda_Mueller_Leskovsek_Duckworth_2017} is a virtual reality game incorporating deep breathing. The associated study showed that this breathing helped them relax their muscles. Therefore, in our study, we asked participants to perform deep breathing before playing the game to help them relax their muscles. We believe this could help reduce any anxiety or the uncomfortable feeling that initial EMS experiences might cause \cite{Knibbe_Alsmith_Hornbæk_2018}.

In summary, many digital games use the body as a unique input source, while the output is typically relayed through a screen. Although some games started to prompt players to look “inward”, the feedback is still mostly presented externally, drawing focus away from the body. Interestingly, prior work proposed using Electrical Muscle Stimulation (EMS) technology to use the body as input and output, explained in the next section.

\subsection{Using the Body as Output by Using EMS}\label{sec2.3:bodyasoutput}

To see our Body as a Play Material approach come to fruition, we utilised Electrical Muscle Stimulation (EMS) technology. EMS has been employed to engage the body as an input and output medium  \cite{Knibbe_Alsmith_Hornbæk_2018,Lopes_Ion_Mueller_Hoffmann_Jonell_Baudisch_2015}. Users attach electrodes to their skin, and when they loan bodily control to "it", a small amount of electric current passes through the electrodes, contracting their muscles and moving their body parts involuntarily \cite{Knibbe_Alsmith_Hornbæk_2018}. We use the word "loan" because the user can resist the electricity and regain control over their body movements.

The process of positioning the electrodes to actuate specific muscles and create certain body movements is called calibration \cite{Knibbe_Alsmith_Hornbæk_2018}. Research suggests that users can initially feel uncomfortable with the EMS sensations, but this can be reduced by spending time during the calibration process \cite{Knibbe_Alsmith_Hornbæk_2018,Lopes_Ion_Mueller_Hoffmann_Jonell_Baudisch_2015}. Moreover, due to differences in users' muscle composition or changes in their body orientation, calibration can sometimes fall short, causing unexpected and ambiguous modifications to EMS-actuated body movements \cite{Knibbe_Alsmith_Hornbæk_2018}. Based on this knowledge, we provided a detailed description of the calibration process to our participants in our pre-study session.

EMS has been used for many playful applications (e.g., \cite{Auda_Pascher_Schneegass_2019,Lopes_Baudisch_2017,Nisal_Patibanda_2022}), particularly in the form of force feedback to make virtual reality games more immersive by, for example, allowing players to experience their hands move involuntarily when hitting a tennis ball \cite{Farbiz_Yu_Manders_Ahmad_2007} or to experience the pushback when touching a virtual wall \cite{Auda_Pascher_Schneegass_2019,Farbiz_Yu_Manders_Ahmad_2007}. In these applications, the player experiences EMS as complementary to the visual feedback. We note that these systems do not indicate to the users when EMS is about to take control of their body. As we did not want to facilitate a surprising experience with the EMS-controlled hand, we learnt from Benford et al. \cite{Benford_Ramchurn_Marshall_2020} who suggested informing the user when a computer is about to take control. Therefore, we used sound in our games so that the player knows when the EMS hand is about to get actuated.

EMS has also created movement-based games such as “Hot-Hands” \cite{Lopes_Ion_Mueller_Hoffmann_Jonell_Baudisch_2015}, where players can play against themselves, albeit to explore proprioceptive interaction \cite{Lopes_Ion_Mueller_Hoffmann_Jonell_Baudisch_2015}. This paper builds on our initial work that \cite{Patibanda_Li_Chen_Saini_Hill_2021,Patibanda_Van_Mueller_2022} has explored variations of these games, but in the form of a pilot study. Hence, we still do not entirely understand the design and associated user experiences of using the body (as input) to play with the body (as output). Therefore, our work aims to begin filling this gap by answering our research question: \textit{how do we design games that allow a player to focus solely on themselves, where they “use” their body to play “with” the body?}

\section{DESIGNING GAMES THAT USE THE BODY AS A PLAY MATERIAL}\label{sec3:designinggames}

We designed three Auto-Paízo games: “Elements”, “Numbers”, and "Slap-Me-If-You-Can", to showcase that our approach is not restricted to one instance. Moreover, having three games allowed us to gather wider insights into user experiences associated with our approach. Our games are varied according to three previously identified key characteristics of bodily games \cite{Mueller_Isbister_2014}: motor-movement \cite{Mueller_Agamanolis_Picard_2003}, game outcome \cite{Isbister_Mueller_2015} and bodily interference \cite{Mueller_Gibbs_Vetere_Edge_2017} (Table \ref{tab:designkeychars}).

\begin{table}[hbt!]
  \caption{Key characteristics of the three Auto-Paízo games.}
  \label{tab:designkeychars}
  \centering
  \begin{tabularx}{\textwidth}{>{\centering\arraybackslash}X>{\centering\arraybackslash}X>{\centering\arraybackslash}X>{\centering\arraybackslash}X}
    \toprule
    \textbf{Game name} & \textbf{Type of motor-movement} & \textbf{Game outcome} & \textbf{Bodily interference} \\
    \toprule
    Elements & Fine motor-movement & Psychological (Symbolic) & Non-interference \\
    \midrule
    Numbers & Fine motor-movement & Psychological (Numerical) & Non-interference \\
    \midrule
    Slap-Me-If-You-Can & Gross motor-movement & Physiological & Interference \\
    \bottomrule
  \end{tabularx}
\end{table}

\begin{itemize}
  \item \textit{Motor-movement} – does the game control fine- or gross-motor movement?
  \item \textit{Game outcome} – is the game outcome a psychological (symbolic, numerical) or physiological?
  \item \textit{Bodily interference} – does the game support interference play (where the EMS-controlled hands interfere with other body parts while playing) or non-interfering play (where the EMS-controlled hands do not interfere with other body parts while playing)?
\end{itemize}

\begin{figure}[h]
    \centering
    \includegraphics[width=1\textwidth]{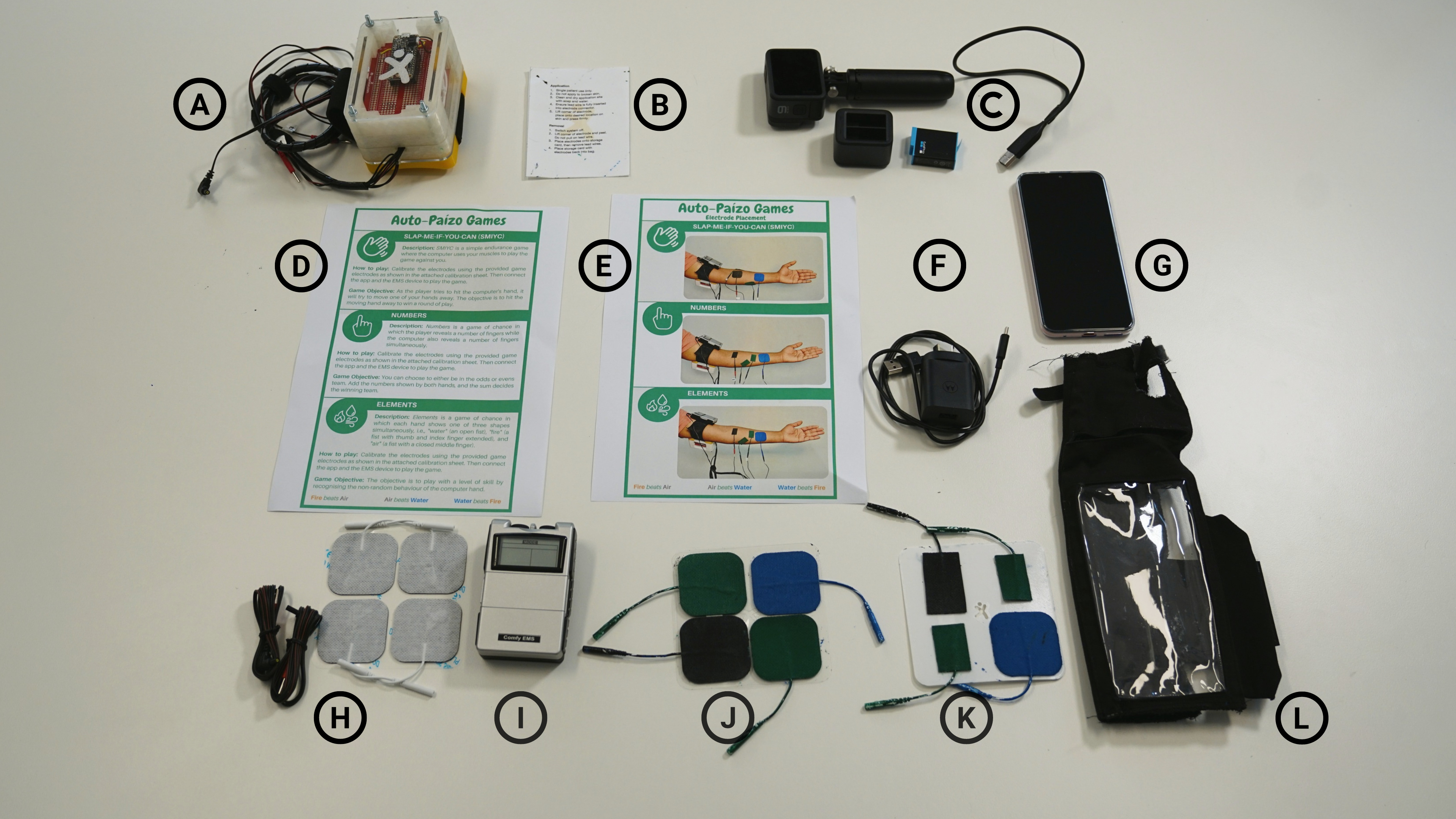}
    \caption{The EMS assemblage is given to each participant. It contains (A) a microcontroller device, (B) EMS instructions, (C) a GoPro action camera with a spare battery and charger, (D) and (E) game descriptions, (F) a mobile charger, (G) mobile phone, (H) spare electrodes and lead cables, (I) EMS device, (J) and (K) electrodes for the games, and (L) hand glove for the mobile phone.}
    \Description{The EMS assemblage is given to each participant. It contains (A) a microcontroller device, (B) EMS instructions, (C) a GoPro action camera with a spare battery and charger, (D) and (E) game descriptions, (F) a mobile charger, (G) mobile phone, (H) spare electrodes and lead cables, (I) EMS device, (J) and (K) electrodes for the games, and (L) hand glove for the mobile phone.}
    \label{fig:participantkit}
\end{figure}

The following sections use the conceptual model of “describing games” \cite{Björk_Holopainen_2003}; first, we describe the hardware and software (together called the “EMS assemblage”) and the gameplay preparation steps that are the same for all games, then we describe each game’s particular design features individually.

\subsection{Game Hardware}\label{sec3.1:gamehardware}

\begin{figure}[h]
    \centering
    \includegraphics[width=1\textwidth]{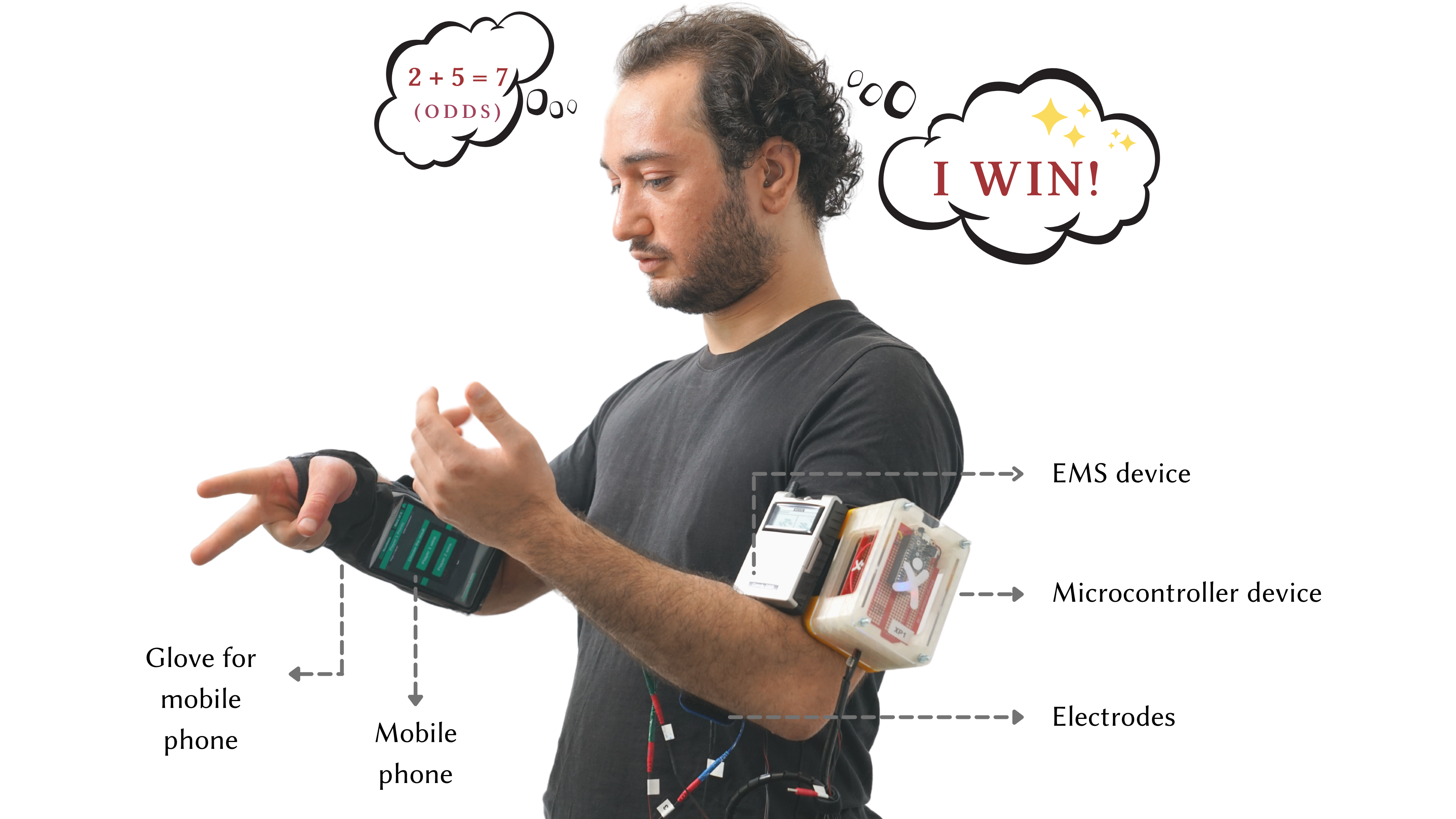}
    \caption{A participant is wearing the hardware to play “Numbers”. The participant chooses their non-EMS right hand to be “odds”, and the EMS left hand to be “evens”. The with their right non-EMS hand shows a “2”, and their left EMS hand shows a “5” simultaneously, making the sum odd (2 + 5 = 7). Therefore, the player’s right non-EMS hand wins.}
    \Description{A participant is wearing the hardware to play “Numbers”. The participant chooses their non-EMS right hand to be “odds”, and the EMS left hand to be “evens”. The with their right non-EMS hand shows a “2”, and their left EMS hand shows a “5” simultaneously, making the sum odd (2 + 5 = 7). Therefore, the player’s right non-EMS hand wins.}
    \label{fig:paizotechsetup}
\end{figure}

\begin{figure}[h]
    \centering
    \includegraphics[width=1\textwidth]{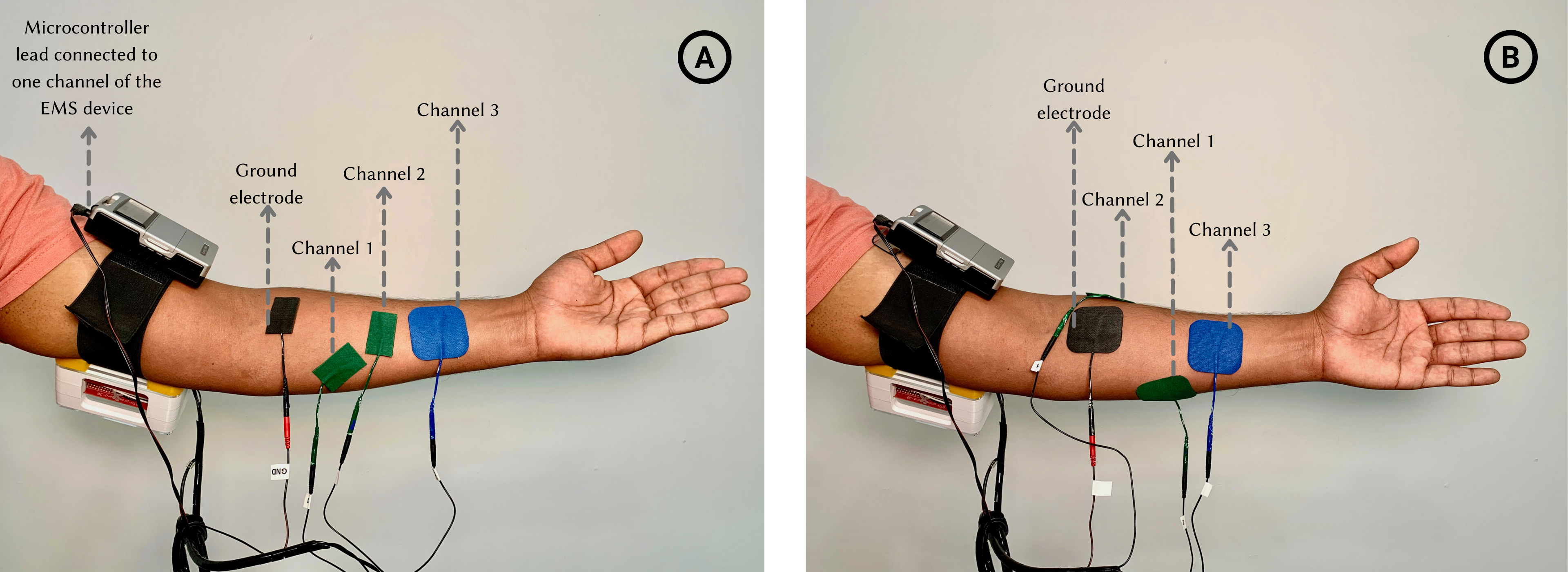}
    \caption{(A) shows the electrodes’ position for Elements and Numbers, and (B) shows the position of the electrodes for Slap-Me-If-You-Can. The size of channels 1 and 2 are smaller in (A) to target muscles for fine motor-movements and larger in (B) to target bigger muscles for gross motor-movements.}
    \Description{(A) shows the electrodes’ position for Elements and Numbers, and (B) shows the position of the electrodes for Slap-Me-If-You-Can. The size of channels 1 and 2 are smaller in (A) to target muscles for fine motor-movements and larger in (B) to target bigger muscles for gross motor-movements.}
    \label{fig:paizotechsetupwithparticipant}
\end{figure}

The hardware consisted of a microcontroller, an EMS device, and electrodes worn on one arm (the EMS hand), while the non-EMS hand has a mobile phone in a custom-made hand glove (Fig. \ref{fig:participantkit}). A commercial EMS device was used to increase players' sense of safety \cite{Knibbe_Alsmith_Hornbæk_2018,Lopes_Ion_Mueller_Hoffmann_Jonell_Baudisch_2015}. The software used the phone's accelerometer sensor to trigger the microcontroller, which actuated the right combination of three electrodes to move the player's EMS-controlled hand involuntarily to play the games. The system was designed to be small and mobile, weighing about 400 grams, with a microcontroller that splits the EMS output into three controllable channels. The microcontroller was placed in a 3-D printed case attached to the arm and the EMS device via a lead. While the hardware was the same for all three games, the electrodes used for each game differed (Fig. \ref{fig:paizotechsetupwithparticipant}). We conducted basic tests such as a drop and shake test and checked how long it takes for the battery to run out (lasted for more than seven days without charging) \cite{Shake}.

\subsection{Game Software}\label{sec3.2:gamesoftware}

\begin{figure}[h]
    \centering
    \includegraphics[width=1\textwidth]{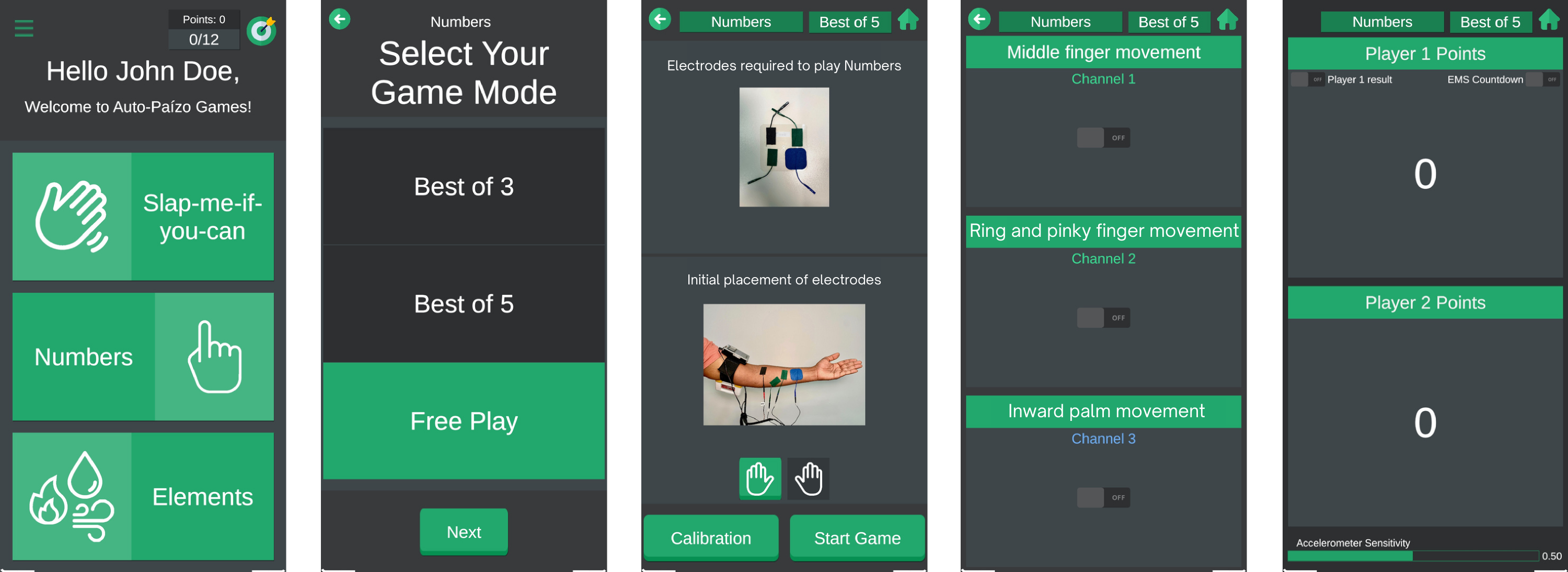}
    \caption{The screens for game and mode selection, calibration instructions, calibration switches and score-keeping.}
    \Description{The screens for game and mode selection, calibration instructions, calibration switches and score-keeping.}
    \label{fig:gamesoftware}
\end{figure}

We developed a phone application that communicates with the microcontroller through Bluetooth. The minimalistic software interface (Fig. \ref{fig:gamesoftware}) was designed to keep players focused on their body movements rather than the screen. The application supports four key actions: logging in, selecting a game and game mode, calibrating the EMS, and tracking the score. Players log the result after each round of play on the score-keeping screen. The application also logs game data, such as the number of times each game was played and features 12 challenges to keep players engaged. We added the challenges considering the games’ simplicity and the study's length to facilitate participants’ engagement with the system as often as possible. The EMS-controlled hand was referred to as "Player 1" and the non-EMS hand as "Player 2" to not influence how players perceived the EMS hand. Unpredictability in the actuation of the EMS hand due to body orientation, as learned from prior work \cite{Knibbe_Alsmith_Hornbæk_2018}, was addressed with the "Player 1 result" feature. When enabled, players could see what the EMS hand meant to show on the screen (not applicable for Slap-Me-If-You-Can). Each gesture was associated with the same sound played for half a second before the EMS hand to help the player know when the EMS is about to take control, while not giving away what gesture the EMS hand would show. An “EMS countdown" feature was designed to facilitate a playful attitude. When enabled, the EMS actuates in sync with the player performing an up-and-down movement countdown (3-2-1) with their non-EMS hand (as known from rock-paper-scissors): the non-EMS hand mirrors the movement by gently curling three times inwards (actuated for one second each time). This movement is complemented by a “dramatic” sound designed to amplify the experience of the countdown movement.

\subsection{Gameplay Preparation}\label{sec3.3:gameplayprep}
Players calibrate the EMS using the software's guidance after attaching the electrodes to their body. The application asks the player to slowly increase the intensity of each channel by turning the dial on the EMS device until they achieve the desired motor-movement. The player might need to reposition an electrode if there is no movement. After repeated testing on the authors' bodies, the pulse rate and width are initially set at 220 ms and 45 Hz. Players are free to adjust these parameters and are encouraged to continue fine-tuning these parameters to achieve the desired movement at a comfortable actuation level. To protect the player from muscle fatigue, we added an alarm in the software that notified the players every 30 minutes. This notification pauses the game (even if the player is in the middle of gameplay) and shows a message on the phone asking players to “rest their body before playing more”. We explain each game in the following sections.

\subsection{Game 1: Elements}\label{sec3.4:elements}
\begin{figure}[hbt!]
    \centering
    \includegraphics[width=1\textwidth]{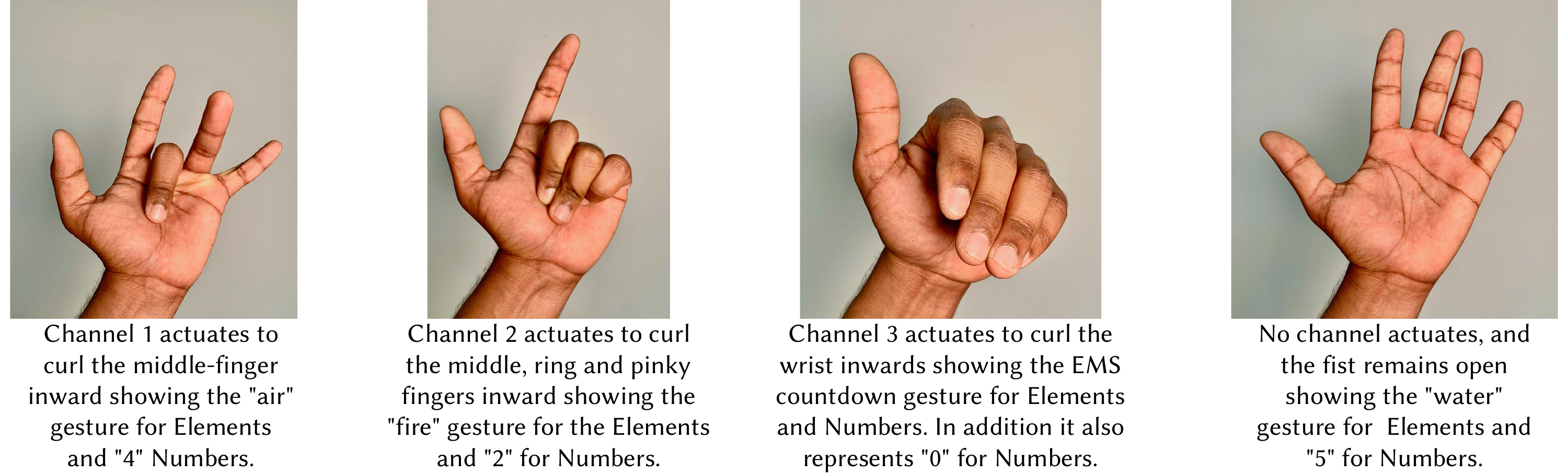}
    \caption{The three EMS channels that show the gestures Elements and Numbers game and their associated meanings.}
    \Description{The three EMS channels that show the gestures Elements and Numbers game and their associated meanings.}
    \label{fig:handcurlsmeanings}
\end{figure}

\begin{figure}[hbt!]
    \centering
    \includegraphics[width=1\textwidth]{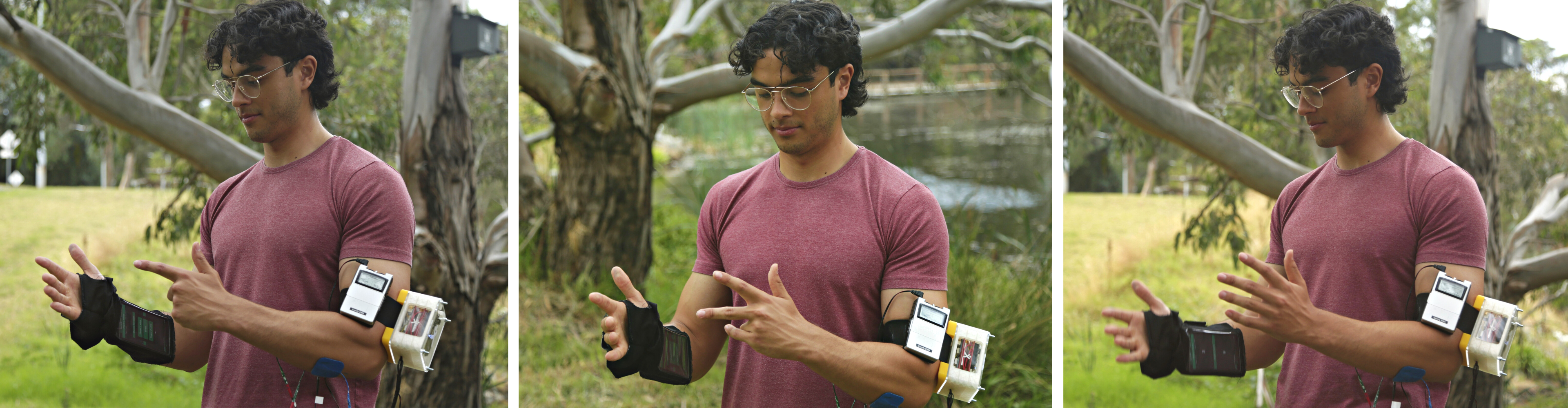}
    \caption{A participant is playing a best-of-three rounds game of Elements.}
    \Description{A participant is playing a best-of-three rounds game of Elements.}
    \label{fig:Elements}
\end{figure}

In the Elements game (Fig. \ref{fig:Elements}), players perform one of three gestures: “fire”, “water”, or “air” (Fig. \ref{fig:handcurlsmeanings}). The rules are: “water” beats “fire”, “fire” beats “air”, and “air” beats “water”. The goal is to show an element that beats the other hand. The game starts with players initiating the countdown by moving their non-EMS hand up and down three times. Then both hands show a gesture simultaneously to determine a winner. Fig. \ref{fig:Elements}’s left image shows the non-EMS hand with “water” and the EMS hand with “fire”, giving the non-EMS hand one point. In the central image, the non-EMS hand shows “water”, and the EMS hand shows “air” (EMS-hand gets the point). In the right image, the non-EMS hand shows “water” again while the EMS hand shows “water”, awarding one point each, making the round a draw.

\subsection{Game 2: Numbers}\label{sec3.5:Numbers}
\begin{figure}[hbt!]
    \centering
    \includegraphics[width=1\textwidth]{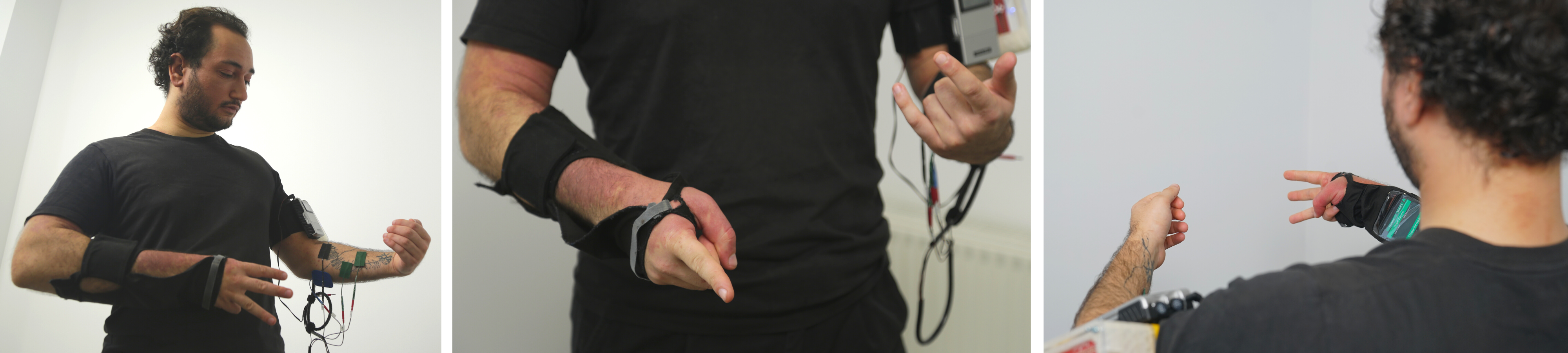}
    \caption{A participant is playing a best-of-three rounds game of Numbers.}
    \Description{A participant is playing a best-of-three rounds game of Numbers.}
    \label{fig:Numbers}
\end{figure}

The Numbers game starts with the EMS hand in an open-fist position and the non-EMS hand in a closed fist position. The player mentally chooses whether they are in the “odds” or “evens” team. After the countdown, both hands reveal a number simultaneously, and the sum determines the winner. For example, in Fig. \ref{fig:Numbers}, a player chooses to be "odds". The left image (round 1) shows the player's non-EMS hand revealing a "3" and the EMS hand showing a "0". The sum is odd (3+0=3), giving the player's non-EMS hand one point. The central image (round 2) shows the player’s hand with a “1” and the EMS hand with a “3”. The sum is “4”, giving the EMS hand one point. The right image (round 3) shows the player’s non-EMS hand with a “3” and the EMS hand with a “5”. The sum is “8”. This result gives the EMS hand one point, making “it” the game-winner.

\subsection{Game 3: Slap-Me-If-You-Can}\label{sec3.6:SMIYC}

\begin{figure}[hbt!]
    \centering
    \includegraphics[width=1\textwidth]{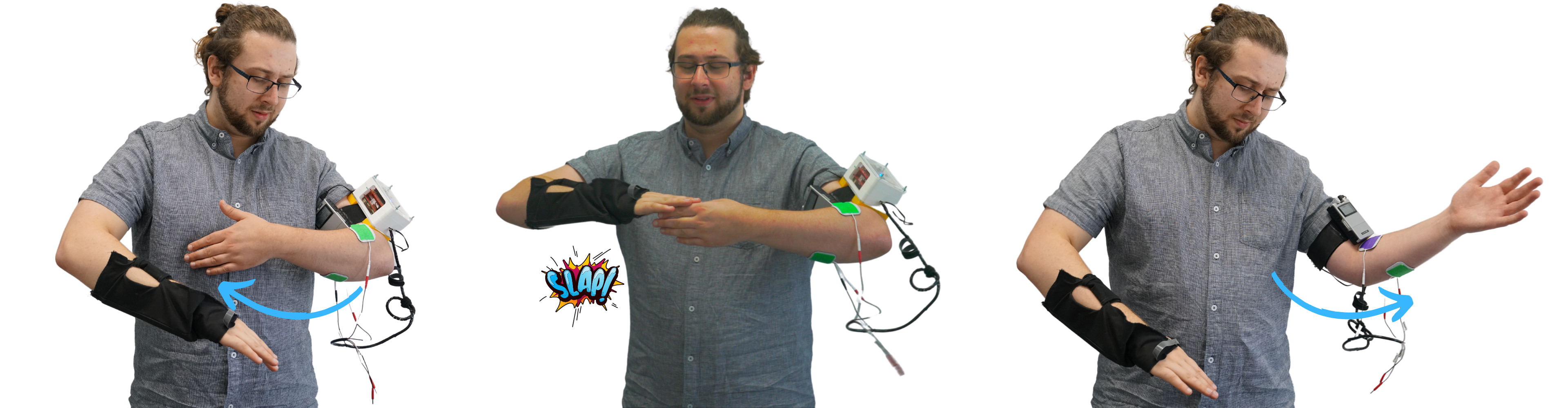}
    \caption{A participant is playing a best-of-three rounds game of Slap-Me-If-You-Can.}
    \Description{A participant is playing a best-of-three rounds game of Slap-Me-If-You-Can.}
    \label{fig:SMIYC}
\end{figure}

Slap-Me-If-You-Can (Fig. \ref{fig:SMIYC}) is inspired by Red-hands \cite{twoplayergamesHands}. The player starts with the non-EMS hand at a 90-degree angle, with fingers facing upward, and the EMS hand is positioned towards the torso with the palm facing upward. After the countdown, the player tries to slap the EMS hand quickly. Successful slaps give one point to the player, while misses give one point to the EMS hand. In Fig. \ref{fig:SMIYC}’s left image, the EMS hand dodges the slap by moving inwards, giving the EMS hand one point. In the central image, the player's hand successfully slaps the EMS hand, giving the player's hand one point. The right image shows the EMS hand avoiding the slap by moving outwards, giving the EMS hand one point and making “it” the game-winner for this best-of-three-round.

\section{STUDY DESIGN AND DATA ANALYSIS}\label{sec4:studydesign}
\begin{figure}[hbt!]
    \centering
    \includegraphics[width=1\textwidth]{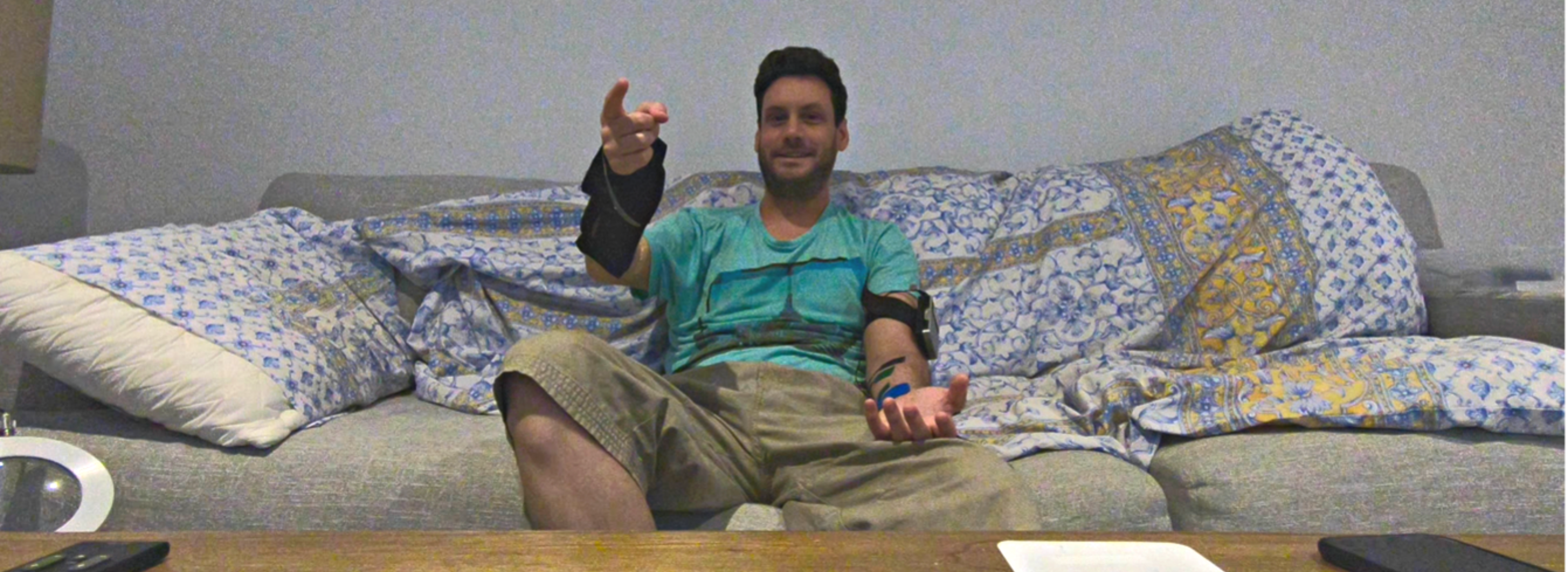}
    \caption{A participant playing the Numbers game before bedtime by relaxing the EMS hand (left) on their thigh.}
    \Description{A participant playing the Numbers game before bedtime by relaxing the EMS hand (left) on their thigh.}
    \label{fig:beforebedtime}
\end{figure}

To understand the user experience of interacting with our games, we obtained institutional ethics approval to conduct a field study \cite{Grzyb_Dolinski_2021}. As shown in previous research on novel game experiences \cite{Semertzidis_Scary_Andres_Dwivedi_Kulwe_Zambetta_Mueller_2020}, this approach allows us to collect rich data from participants, whilst reducing potential influences of the researchers’ personal biases. This method proved to be particularly beneficial during the COVID-19 pandemic when our study was conducted. It facilitated remote engagement with our system, allowing participants to play the games from their homes, at their convenience, and without researchers' immediate presence. It further allowed participants to explore the system in their own ways, such as creating their own games (Fig. \ref{fig:socialEMSgames}) or using the games as a relaxation tool before bed (Fig. \ref{fig:beforebedtime}). These instances suggest the potential for this method to capture participants' preferences in system use, rather than adhering strictly to predetermined expectations by researchers \cite{Grzyb_Dolinski_2021}. To engage players over these seven days, we also provided them with game challenges (Appendix A: Table. \ref{tab:gamechallenges}).

\begin{figure}[hbt!]
    \centering
    \includegraphics[width=1\textwidth]{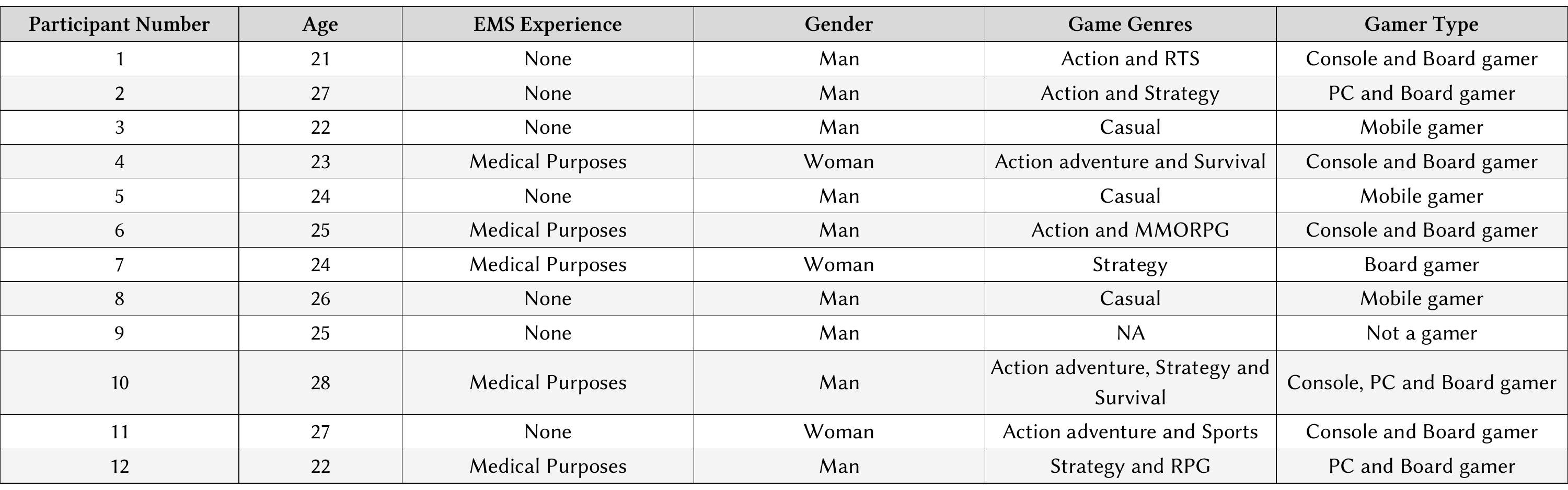}
    \caption{Demographics including gender distribution, EMS experience, preferred game genres, and gamer types.}
    \Description{Demographics including gender distribution, EMS experience, preferred game genres, and gamer types.}
    \label{fig:demographics}
\end{figure}

The study had three phases: a pre-study phase, a field study phase, and a post-study interview phase. We recruited 12 participants by advertising through our social media channels and mailing lists. Out of the 12 participants, 75\% (n=9) identified as men and 25\% (n=3) as women (none non-binary or self-described). The participants' mean age was 24.5 years (SD=2.23). Detailed demographics are in Fig. \ref{fig:demographics}.

\subsection{Pre-study Phase (Safe Onboarding and Informing System’s Capability)}\label{sec4.1:prestudy}
We delivered the EMS assemblage to each participant (adhering to COVID-19 restrictions) before setting up a video conference call. The goal of the pre-study phase was to educate participants about the "dos and don'ts" of using an EMS device and relax their body muscles by breathing before calibrating their hands to play all three games at least once. During this phase, participants were also provided with the research team’s contact details and informed that they could contact us during working hours to inform or rectify any hardware, software or study-related issue. While one participant could not calibrate for any game, they retained the EMS assemblage during the study period. They attempted to calibrate and play without the research team’s support during the field study phase. Consequently, we report on the experiences of 11 participants playing the games and 12 using the system. Also, one participant could not calibrate for Elements and hence only played Slap-Me-If-You-Can and Numbers for the study phase. The average time for the pre-study phase was 58 minutes, with a standard deviation of 18 minutes.

\subsection{Field Study Phase (Feedback, Study and Data Collection)}\label{sec4.2:fieldstudy}
Participants had the EMS assemblage for seven days and were asked to play each game at least seven times. Participants were also asked to record themselves while playing using the provided GoPro action camera. Based on our log data, we analysed that each participant played “Elements”, “Numbers”, and “Slap-Me-If-You-Can” for an average of 15, 13, and 8 times, respectively. This means that, on average, each participant engaged with the games for (15+13+8) * 3 (average time to play each game), making it a total of 108 minutes over the seven days. This number does not include the calibration time.

\subsection{Post-study Phase (Further Data Collection and Data Analysis)}\label{sec4.3:poststudy}
Following the field study phase, we conducted interviews with participants, lasting for a mean duration of 59.91 minutes (standard deviation = 22.40 minutes), to gain insights into their experiences. These audio and video recorded interviews employed a semi-structured approach \cite{Dearnley_2005}, providing flexibility in capturing a rich set of subjective responses. The interviews were then transcribed, after which two coders separately conducted an inductive thematic analysis \cite{Braun_Clarke_2006} using NVivo software \cite{Wiltshier_2011}. We chose the inductive thematic analysis method for its ability to distil and articulate patterns within our dataset, allowing the data to drive the identification of our themes. This method helped us explore individual experiences and recurring patterns across our participants. When articulating our themes, our goal was not to make broad generalisations but to explore the various nuanced aspects of our participants’ experiences, extending beyond usability and the novelty of the experience. As such, this method met our aim to understand the novel interaction offered by our games.

We considered each answer (average word count = 58) as one data unit. Each coder independently read all 704 data units two times and assigned every data unit a category code. Coders 1 and 2 created 82 and 75 initial codes for all the data units. The coders then communicated with each other to refine the initial codes down to 64 final codes. The inter-rater reliability was 97.43\%, and Cohen’s kappa coefficient (k) = 0.653. The coding categories were then examined, cross-referenced with the data units, and further analysed for overarching themes, which both researchers again reviewed. Finally, we found seven overarching themes, of which we report on four that align with our research question. These four overarching themes were developed using 41 of the 64 final codes, and these 41 codes comprised 360 of the 704 data units.

\section{QUANTITATIVE RESULTS: ENGAGEMENT PATTERNS FOR ELEMENTS, NUMBERS, AND SLAP-ME-IF-YOU-CAN GAMES}\label{sec5:quantitativeresults}
A comprehensive analysis of the engagement patterns across the three games, Elements, Numbers, and Slap-Me-If-You-Can, revealed nuanced similarities and differences in the playing behaviour over the seven days.

\begin{itemize}
    \item \textbf{The total number of games played:} The total number of games played showed that Elements recorded the highest engagement with 179 games played (14.92 games/participant, SD = 5.60), followed by Numbers with 157 games (13.08 games/participant, SD = 4.74), and Slap-Me-If-You-Can with 101 games (8.42 games/participant, SD = 3.15), suggesting that the Elements game was the most appealing overall.
    \item \textbf{Daily engagement patterns and variability:} For Elements, engagement peaked on Day 1 with an average of 4.08 games per participant (SD = 3.71) and generally declined over the week, reaching its lowest point on Day 4 with 0.92 games per participant (SD = 1.16). Numbers exhibited the highest engagement on Day 1 at 3.5 games per participant (SD = 2.67). It fluctuated throughout the week and the lowest engagement on Day 3 at 0.83 games per participant (SD = 1.03). With Slap-Me-If-You-Can, engagement was the highest on Day 1 with 2 games per participant (SD = 1.67). It fluctuated throughout the week, reaching its lowest engagement on Day 7 with 0.67 games per participant (SD = 0.98). These patterns suggest that the Elements game's novelty may have worn off over time, while Numbers and Slap-Me-If-You-Can maintained a more lasting appeal for some participants.
    \item \textbf{EMS experience and game engagement:} We observed several patterns when considering prior EMS experience and game engagement. With Elements, participants with medical experience had an average engagement of 2.67 games/participant (SD = 2.31), and those with research experience had an average of 14 games/participant (no SD available as only one participant had research experience). Participants with no EMS experience had an average engagement of 2.5 games/participant (SD = 1.58), with the single participant using EMS for research purposes having the highest engagement. With Numbers, participants with no experience had the highest average engagement (average 3.5 games/participant; SD = 1.73), followed by the participant with research experience (2.83 games/participant; no SD available) and those with medical experience (average 2.67 games/participant; SD = 1.25). Similarly, with Slap-Me-If-You-Can, participants with no experience had the highest average engagement (average 1.83 games/participant; SD = 0.83), followed by the participant with research experience (1.5 games/participant; no SD available) and those with medical experience (average 1.33 games/participant; SD = 0.82).
\end{itemize}

In summary, Elements was found to have higher engagement levels for participants with prior EMS experience for medical and research purposes. In comparison, Numbers and Slap-Me-If-You-Can had higher engagement levels for participants with no previous EMS experience. Day 1 had the highest engagement levels for all three games, but Elements consistently declined throughout the week. In contrast, Numbers Slap-Me-If-You-Can games experienced more fluctuations.

\section{QUALITATIVE RESULTS: FOUR PLAYER EXPERIENCE THEMES}\label{sec6:themes}
At the beginning of the study, some participants expressed their apprehension about using EMS. However, they enjoyed playing the Auto-Paízo games over the study period. For example, P4 said, \textit{“It was scary at first, but it felt nice over time, in a weird way”}. In this section, we articulate and discuss four overarching themes that describe the participants’ experiences: 1) loaning bodily control to a computer, 2) playfully engaging with the integrated computer, 3) playing with the ambiguity of the computer-controlled body parts, and 4) bodily awareness when loaning bodily control to a computer to play.

\subsection{Theme 1: Loaning bodily control to a computer}\label{sec6.1:theme1}
In this theme, our participants discussed how they allowed EMS to control their hand movements and how the speed of the EMS hand’s movement affected their experience of playing with a computer-controlled version of themselves. Eleven of the 41 codes are associated with this theme, representing 84 data units. The theme has two sub-themes: learning to adapt to the body to turn it into a playful material (representing six codes and 45 data units), and influence of the EMS hand’s movement speed on gameplay (five codes and 39 data units).

\subsubsection{Learning to Adapt to the Body to Turn It Into a Playful Material}\label{sec6.1.1:subtheme1}
Eleven of our twelve participants discussed their experiences when they were learning how to give away bodily control to the computer to turn it into a playful material. Participant P7 described the relationship they tried to have with the EMS and said, \textit{“It is my body, but I am trying to use it to listen to [the] computer.”} When questioned about how this “listening” worked, P7 said, \textit{“I relax my muscles for the computer [to] work well so I can play with it.”} Here, the participant relaxed their muscles to allow the EMS to work best. When participants did not relax, the EMS could not always actuate their muscles fully, so the gestures could be incomplete. For example, P3 said, \textit{“Sometimes, I just forgot to relax my EMS hand, leading to incomplete gestures.”} As such, participants said that they were actively involved when allowing the computer to use their body. For example, P8 said, \textit{“I tried to increase the intensity of the EMS and let it control my body.”} P3 added, \textit{“I initially forgot the advice you [referring to the researcher] gave about breathing. Later, this breathing helped me let go of my EMS hand.”} 

Participants suggested that turning the body into a playful material is a personal experience. P10, who tried to calibrate the system but could not achieve the desired hand movements, said, \textit{“It was fascinating to see that your [referring to the researcher] hand was moving, but I was surprised and sad that my fingers were not moving. I increased the intensity to maximum comfort, but it was still not working.”} Participants who relaxed their muscles could mostly experience complete actuation. However, the others experience incomplete actuation. P9 suggested that such ambiguous body movements were also a source of their enjoyment and said, \textit{“It’s fun to anticipate what the EMS hand tried to show when the actuations are fully executed by the EMS.”} This participant also said, \textit{“When I am anticipating what is going to happen, I sometimes forget that I have to play with my hand too because it is fascinating to watch the hand move involuntarily.”} These participant responses suggest they tried varied ways to turn their body into a playful material. While some sought to change the system’s parameters (EMS intensity), others tried to relax their bodies by breathing (or both).

\subsubsection{Influence of the EMS Hand’s Movement Speed On Gameplay}\label{sec6.1.2:subtheme2}
Participants discussed the influence of the EMS-hand’s movement speed on their ability to loan bodily control. Five participants said that they had trouble playing Slap-Me-If-You-Can. P6 said, \textit{“I was easily slapping “it” most of the time.”} While our participants believed their muscles were actuated promptly, they felt the arm movements were slower for this game. P7 reflected on this matter and said, \textit{“I think bigger movements were required to play Slap-Me-If-You-Can, and my body could not respond to the EMS actuations quickly enough.”} P1, who experimented with the EMS intensity while playing this game, said, \textit{“It gets too uncomfortable too quickly when I increase the intensity.”} We also enquired if participants felt the EMS moved their hands slowly, even during calibration. P6 said, \textit{“No, during calibration, I didn’t feel there was a lag in-between toggling the channels on the application and the response of my EMS hand.”} P9 reflected on this subject and said, \textit{“During calibration, I was only focusing on one hand, and therefore did not have my other hand moving to compare the speed. However, I could notice the speed difference while playing as I needed to control my non-EMS hand to play the game.”}

Four participants reported they enjoyed playing Slap-Me-If-You-Can. P2 said, \textit{“My hand was moving quickly enough.”} P12 believed they added their own force to the EMS hand’s movement and said, \textit{“Maybe because I like physical games, I felt like I supported the computer’s movements with my own [laughs].”}

Participants also compared their experience of movement speed across all three games. P2 said, \textit{“When playing the Elements and Numbers game, you only move your body very slightly, just your fingers. However, you are moving your upper body forward to slap the EMS hand when playing Slap-Me-If-You-Can. Therefore, while thinking I was letting the EMS control my hand, I might have taken back control without your knowledge. This could have caused the delay because you are restricting the computer from acting.”} They also compared the actuation time, and P3 said, \textit{“The short actuation time was enough to control small finger movements in Numbers and Elements and was not enough for the bigger movements required for Slap-Me-If-You-Can. However, I could not control this parameter.”}

These insights indicate that participants' engagement in such games can dynamically influence their gameplay experience. Participants can potentially enhance their experience by intentionally loaning more control of their body to the computer. Conversely, they could diminish their gameplay experience by obstructing the computer's control. This might occur, for example, if participants focus too heavily on the computer's movements rather than on the body parts they themselves are controlling.

\subsection{Theme 2: Playfully engaging with the integrated computer}\label{sec6.2:theme2}
This theme discusses how participants used the games’ varieties to playfully engage with the integrated computer. Nine of the 41 codes were associated with this theme, representing 90 data units. This theme has two sub-themes: varied attitudes towards the computer-controlled hand when playing the games (five codes and 53 data units) and various ways to engage with the body during gameplay (four codes and 37 data units).

\subsubsection{Varied Attitudes Towards the Computer-controlled Hand when Playing the Games}\label{sec6.2.1:subtheme1}
Nine participants discussed their attitude towards the computer-controlled hand. P6 said, \textit{“I treated the EMS hand as a different person, but it was still mine.”} Participants discussed the competition and game outcome and noted, \textit{“It was weird to experience losing to myself if the EMS hand won”} (P6). Some participants spoke to the EMS-controlled body part, with P1 saying, \textit{“I personalised the experience by speaking to it, and after every result, I was trying to speak to myself and ‘it’ to feel the presence of another player.”} P8 expressed a sense of social play, saying, \textit{“If I am playing against someone, there will be a psychological banter, but here I am playing against a robot. So, I felt the fun you achieve is not so much because you are still playing to compete against yourself.”} On the other hand, some participants (n=5) did not focus on the competition but on interacting with the EMS hand. P3 intended \textit{“to play and interact with the EMS hand.”} They did not care about the result because \textit{“I knew I was not playing with somebody else.”} Similarly, P4 said, \textit{“My goal was to have fun while exploring my body”} as the traditional understanding of competition did not apply.

These results show that some participants saw the EMS hand as a personal entity and personalised the experience by speaking to the EMS-controlled body part to foster a sense of social play. Moreover, while some wanted to compete and experienced weirdness when losing to the EMS hand, others did not care, indicating their intention to engage with their body.

\subsubsection{Various Ways to Engage with the Body During Gameplay}\label{sec6.2.2:subtheme2}
Seven participants reflected on the various ways in which they engaged their body to play the games. P3 said, \textit{“I played all the games every day for about 30 minutes but changed the order of these games.”} Four participants said they had a favourite game initially. P4 said, \textit{“Initially, I liked the Numbers game better because I had to calculate and decide the winner.”} Upon enquiring the reason behind this change of preference, they said, \textit{“I felt the result was more direct with the Elements game, meaning I did not have to do any math mentally, and I just understood the result by feeling and looking at my hands compared to the Numbers game.”} Six participants also described how they did not like that they had to keep changing the electrodes to continue playing. P3 said, \textit{“I liked to play the games, but I did not like the associated labour job of changing electrodes.”} Moreover, five participants reflected on the gestures themselves. P5 said, \textit{“I was more comfortable with the EMS hand actuating the "Fire" gesture in the Elements game than the "Air" gesture, as I am not used to just closing my middle finger.”}

\begin{figure}[hbt!]
    \centering
    \includegraphics[width=1\textwidth]{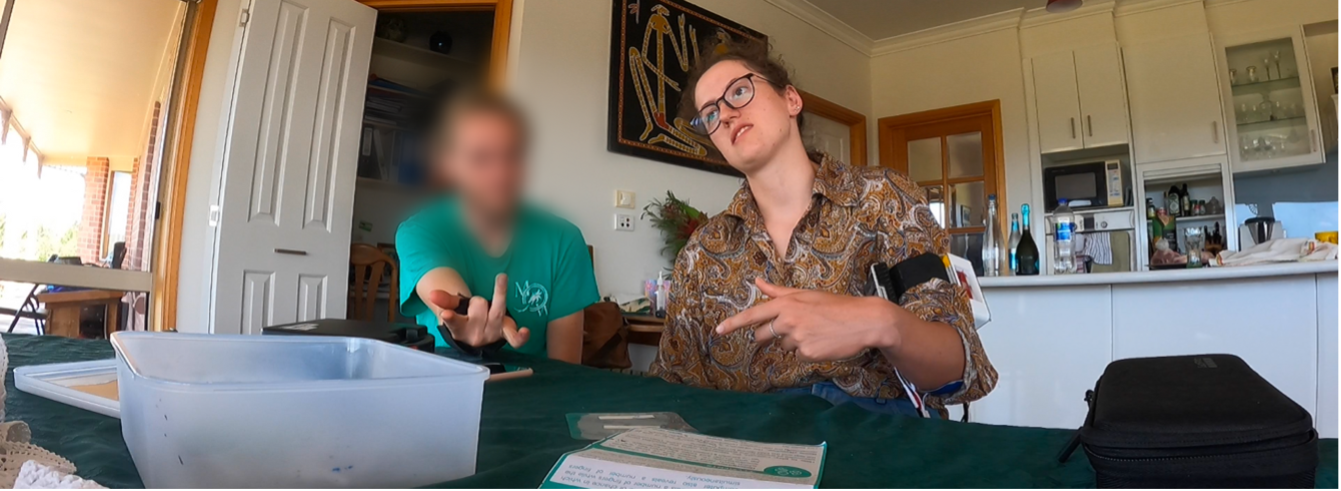}
    \caption{A participant who designed their own social EMS game.}
    \Description{A participant who designed their own social EMS game.}
    \label{fig:socialEMSgames}
\end{figure}

Five of the seven participants also discussed how much control over their body they had to give up playing the three games. P5 said, \textit{“I had to give away only a small amount of bodily control when playing Numbers and Elements, which I was happy about, but a greater amount when playing Slap-Me-If-You-Can.”} P12 added to this commentary, saying, \textit{“Playing Slap-Me-If-You-Can made it more challenging to give away bodily control, and it took a while for me to understand how much I had to relax my muscles so that I could properly play the game.”} Participants also used the EMS system beyond the designed games. Three participants decided to design a game of their own, \textit{“After I finished playing these games, I was playing a prediction game with my boyfriend in which he was wearing the glove with the phone and playing Numbers. His movement, of course, was controlled by the EMS hand, but I was also showing a number with my non-EMS hand. The game was to predict the number the EMS might show.”} (P7) (Fig. \ref{fig:socialEMSgames}).

\subsection{Theme 3: Playing with the ambiguity of the computer-controlled body parts}\label{sec6.3:theme3}
In this theme, participants discussed their experiences dealing with the sometimes-ambiguous movements of the EMS hand and how this led to fun associated with prediction. Nine of the 41 codes are related to this theme, representing 81 data units. This theme has two sub-themes: dealing with the incomplete movements of the computer-controlled hand (4 codes and 35 data units) and predicting the computer-controlled hand’s gameplay (5 codes and 46 data units).

\subsubsection{Dealing with the Incomplete Movements of the Computer-controlled Hand}\label{sec6.3.1:subtheme1}
Eight participants described how the EMS hand occasionally did not finish its movements, with P12 saying that \textit{“sometimes the finger movement was vague.”} P12 continued saying, \textit{“The computer cannot know what it nor I showed while playing the games. I had to be true to play the game.”} P5 added, \textit{“There is the temptation not to be ethical when playing against yourself,”} but tried to resist this. Some participants admitted using this flaw to their advantage in competitive situations. P9 said, \textit{“When I really wanted to win, I could.”}

Five participants discussed determining the outcome when the EMS hand did not fully close its fingers. P7 said that when playing Numbers, \textit{“I knew the EMS hand was showing a three because I could feel what EMS channel got actuated for that number on the skin, but the visual was a two as the third finger didn’t move completely.”} P4 described their response to this situation: \textit{“In these scenarios, I gave the point to the computer.”} However, some participants were unsure whether the computer moved their fingers completely. P3 said, \textit{“Sometimes I was unsure if the computer moved my body or was it just me, moving for the computer.”} Four participants tried to resolve this ambiguity by closing their eyes while playing to improve their perception of what the incomplete hand movements showed. P12 said, \textit{“While playing Elements and when my eyes were open, it was a bit more ambiguous because I was just watching my hand and trying to see what it was doing.”} Switching the hardware between arms was another approach used to reduce ambiguity. P5 said, \textit{“I don’t know why my left hand was not working; therefore, I calibrated my right hand to play the games.”}

\subsubsection{Predicting the Computer-controlled Hand’s Gameplay}\label{sec6.3.2:subtheme2}
Three participants described played with the computer’s ambiguity by trying to predict “its” movements. P2 said, \textit{“I wanted to get better at dealing with the ambiguity initially. But when I didn’t notice any pattern of when the EMS was doing incomplete movements, I tried to predict when it could happen and tried to predict what it was trying to say.”} P4 said, \textit{“Sometimes, the EMS hand’s fingers would not close fully while playing Numbers, and I would just nudge the computer to do the action fully.”}  P5, who used this tactic of prediction in all the games, said, \textit{“I realised that these uncertain/incomplete movements were also fun and tried to do it. However, prediction in the psychological games was more fun than the physiological game.”} 

P1 described how they played such a prediction game and said, \textit{“I would try to predict the EMS hand’s movement by trying to feel the point of actuation on my muscles.”} Upon questioning the reason behind this action, they added, \textit{“When the actuation happened, sometimes, I subconsciously knew what it meant.”} Participants described the effect of the sound feedback coming from the mobile phone, which influenced their experience with this prediction. P1 said, \textit{“I have a hearing problem, and therefore, I did not listen to the sound created right before actuation.”} P3, who did not have any prior EMS experience, said, \textit{“I am more focused on predicting just by focusing on one sense, I could not pay attention to two kinds of feedback at the same time.”} P7 liked the generic said, \textit{“I enjoyed receiving the sound feedback as it helped me give away more control to the computer when it was about to get actuated. I also liked that each gesture has the same sound as it felt like the computer was trying to confuse me and give away the gesture it was about to perform at the same time.”} Participants also utilised the ambiguity arising from the EMS to either support it (work as a team) or use it to their advantage. P12 added to this narrative, \textit{“Sometimes I knew what the EMS wanted to know as I was feeling the actuation on my skin. So, if the actuation didn’t fully happen, I would comply and complete the action the computer was trying to achieve.”}

These results suggest that participants used the ambiguity of the EMS to make gameplay more playful, with some trying to predict incomplete movements to gain an advantage and others working as a team with the EMS hand to support it. One key lesson learned is that users may find ways to deal with incomplete computer-controlled bodily actions, making the experience more engaging and enjoyable. Another lesson is that a multisensory approach could improve the accuracy of predictions and help users deal with ambiguity.

\subsection{Theme 4: Bodily awareness when loaning bodily control to a computer to play}\label{sec6.4:theme4}
In this theme, participants discussed their experiences about how loaning bodily control to a computer to play the games allowed them to reflect and become more aware of their body. 13 of 41 codes were associated with this theme, representing 135 data units. The theme has two sub-themes: reflecting on one’s body due to the computer-controlled hand’s performance (7 codes and 77 data units), and leveraging sensory cues and rhythmic movements to reflect on the body (6 codes and 58 data units).

\subsubsection{Reflecting on One’s Body Due to the Computer-controlled Hand’s Performance}\label{sec6.4.1:subtheme1}
Six participants reflected on how playing reminded them of their past bodily activities while playing the games. P7 said, \textit{“Oh! I think I skied too much last week, and my muscle must be tired.”} Upon asking how they felt remembering this memory, they said, \textit{“It was strange that this game made me recollect that memory. We did have a good time during the Ski trip.”} P9, for whom the actuation did not work for the games for a couple of days in a row, said, \textit{“I was a little concerned as I was not sure why it was not working. I started to think about what I did or did not do over the last two days and realised I was overworked at the gym.”} Two participants specifically played these games after their work time, in the evening. P4 explained their choice and said they like to play these games before bedtime and \textit{“it does make you take 15 minutes out of your day to think about your body, and that’s an interesting concept.”} Upon questioning how they think about their body, P4 said, \textit{“I thought of these games as almost like, mindfulness or meditation activities, where I had only to use one part of my body actively and let the computer play with the other part, I didn’t have to think about this other part and just let it loose.”}

\subsubsection{Leveraging Sensory Cues and Rhythmic Movements to Reflect on the Body}\label{sec6.4.2:subtheme2}
Seven participants described their experience of reflecting on their body with the help of sensory cues and actuated rhythmic movements. P8 specifically discussed how they used their body differently while preparing to play and said, \textit{“When I felt the EMS hand performing the countdown movement for the first time, my immediate reaction was to take back control.”} Upon questioning what this reflection meant to them, P8 added, \textit{“I knew it would take a while to learn to allow the EMS to control my body, but I was unsure of how to do this as I never had to let go of my bodily control consciously.”} P7 said, \textit{“I used the sound of the rhythmic movement and chose to focus on it.”} Upon asking how focusing on the sound helped, P7 said, \textit{“It helped me keep track of my movement count. Since I was counting the countdown, I focused less on the EMS hand, which enabled me to not think of it. It was a tricky learning curve.”} P12 commented on what they liked about this countdown feature and said, \textit{“I liked the feeling of response by the EMS hand. It helped me prepare and know when the EMS will take control of my body.”} P6 added to this topic and said, \textit{“It feels like the EMS has a bit of character and made the gameplay dramatic.”} Overall, participants described that the sound \textit{“added pop to the games”} (P12 and P5), giving the game an \textit{“arcade”} (P6 and P9) feel.

\section{DISCUSSING THE PLAYER EXPERIENCE THEMES}\label{sec7:discussion}
The four themes discussed above describe participants’ experience of engaging with their body as a play material, which can be helpful as descriptive tools for game design researchers. These themes uncovered that our participants appreciated engaging with the variety of bodily movements of the computer-controlled hand for play and the ambiguity of using their body as a play material. Prior work has highlighted that varied bodily movements \cite{Marquez_Segura_2013} and ambiguity \cite{Gaver_Beaver_Benford_2003,Sutton-Smith_2001} are key concepts when designing bodily games \cite{Marquez_Segura_2013,Mueller_Isbister_2014}. We extend these concepts by articulating prescriptive design considerations for our Body as Play Material approach.

\subsection{The Body as a Varying Play Material}\label{sec7.1:bodyasvaryingplaymaterial}
Prior work stressed that variety is an important ingredient for engaging gameplay \cite{Salen_Zimmerman_2004,Waters_Maynard_2014}. Our results confirmed this prior theory and extended it to the Body as Play Material approach. In particular, our participants highlighted that variety in computationally controlled bodily movements could contribute to an engaging gameplay experience. We now discuss four aspects of our games that promoted this variety and turn them into design considerations for designers aiming to utilise this knowledge in their games.

\subsubsection{Consistent Hardware Set Up Across Multiple Simple Games for Deeper Bodily Reflection}\label{sec7.1.1:consistenthardware}
Our game hardware required participants to retain the size and position of the EMS electrodes to play Elements and Numbers but not for Slap-Me-If-You-Can, allowing them to enjoy playing Numbers and Elements more than Slap-Me-If-You-Can (Theme \ref{sec6.1:theme1} and \ref{sec6.3:theme3}). Reflecting on this observation, we note that participants could play two games (Numbers and Elements – as they had a consistent electrode set up) consecutively while having to pause to recalibrate for playing Slap-Me-If-You-Can. This continuity in gameplay might have given them extra time to get accustomed to the EMS actuations, allowing them to achieve a flow state \cite{Csikszentmihalyi_Csikzentmihaly_1990}. Moreover, continued playtime without changing hardware possibly helped players to focus inward and reflect on their own body, especially considering the simplicity of the games. This inward focus might have also helped how the computer would control their body.

In prior work, Mueller et al. \cite{Mueller_Byrne_Andres_Patibanda_2018} highlighted that bodily play experiences should help a human look inward into their living body, not just use the body as an object. Our design supports this theory and extends it to the Body as a Play Material approach by suggesting that designers consider retaining any hardware for controlling the body across multiple simple games for deeper bodily reflection.

\subsubsection{Leveraging the Limitations of the Body-technology Integration to Create Meta-gameplay}\label{sec7.1.2:levraginglimitations}
Game designers often create playful experiences using novel technologies, pushing their boundaries  \cite{Dyck_Pinelle_Brown_Gutwin_2003}. One of the significant limitations of EMS, or so researchers thought \cite{Coyle_Moore_Kristensson_Fletcher_Blackwell_2012,Knibbe_Alsmith_Hornbæk_2018}, is that there is a slight delay (less than half a second) for the electricity passing through the body to transform into actuations. We thought this might also be a limitation. However, our results in Theme 2 suggest that participants used it to their advantage. This temporal gap between the time of actuation and the result getting visually displayed through the actuation on their hand allowed them to play by trying to anticipate the EMS hand’s gesture. Therefore, a player could anticipate by feeling the electricity and waiting for the visual confirmation, allowing them to create a meta-guessing game.

Prior work suggested that the limitations of technology can be used to amplify the player’s experience of focusing on their bodily actions while playing movement-based games \cite{Patibanda_Mueller_Leskovsek_Duckworth_2017}. We support this theory with our work and extend it to the Body as a Play Material approach by suggesting that designers leverage the limitations of the body-technology integration to create meta-gameplay.

\subsubsection{Compensating for the Tediousness of Calibration Through a Minimalist Rule Set}\label{sec7.1.3:compensatetediousness}
As described in section \ref{sec4:studydesign}, the average pre-study took about one hour per participant, out of which a large portion was dedicated to the calibration process (approximately 30 minutes on average), indicating the tediousness of the process. Our results suggest that participants took much less time during the field study phase (15 minutes on average) but indicated that recalibration to change games could still be improved. While researchers are exploring using artificial intelligence to optimise this process \cite{Gange_Knibbe_2021,Knibbe_Freire_Koelle_Strohmeier_2021}, our results suggested that designers could compensate for the tediousness of calibration through a minimalist rule set.

Prior work showed that minimalist games can have a small rule set while still being sufficiently deep \cite{Nealen_Saltsman_Boxerman_2011}. Our work extends this to the Body as a Play Material approach by suggesting that designers could compensate for the tediousness of calibration through a minimal rule set.

\subsubsection{Varying Computer-controlled Movements to Support Reflection on Bodily Limitations}\label{sec7.1.4:varyingmovements}
We used one of the forearms to create fine- and gross-motor movements as this is the most used body part when using EMS in HCI \cite{Faltaous_Koelle_Schneegass_2022}. We designed games in which the EMS and non-EMS hands perform both kinds of gestures to play. Participants reported enjoyment with the variety of movements but expressed discomfort with some unnatural gestures, especially when these gestures were performed by the EMS-controlled hand. For instance, participants felt more at ease with the "Fire" gesture than the "Air" gesture in the Elements game, due to the unnatural motion of solely closing their middle finger. Despite the discomfort, participants found reflecting on and understanding their body movement limitations useful.

Prior work has detailed the importance of combining fine- and gross-motor movements when creating bodily games \cite{Mueller_Agamanolis_Picard_2003} and suggested that supporting varied bodily actions can facilitate diverse bodily reflections \cite{Benford_Greenhalgh_Giannachi_Walker_Marshall_Rodden_2012}. Our work supports this theory and extends it to the Body as a Play Material approach by suggesting that designers consider varied computer-controlled movements to support reflection on bodily limitation \cite{Benford_Greenhalgh_Giannachi_Walker_Marshall_Rodden_2012}.

\subsection{The Body as an Ambiguous Play Material}\label{sec7.2:ambiguousplaymaterial}
Research highlights the importance of ambiguity for play as it can foster a sense of “curiosity, exploration and empowers the player” \cite{Sutton-Smith_2001}. Through our study, we learned that participants elicited various ways to engage and deal with the ambiguity of a computer-controlled body part. This section discusses three aspects of ambiguity that our participants experienced.

\subsubsection{Incorporate Sound to Notify Players when the Computer is About to Take Bodily Control Without Specifying the Nature of this Control}\label{sec7.2.1:incorporatesound}
Designing meaningful play requires players to receive immediate and clear feedback \cite{Salen_Zimmerman_2004}. In our games, since the "other player" was part of their own body, we designed sound feedback so that players knew when the computer was about to take bodily control. Moreover, we played the same sound for every kind of gesture made by the EMS hand to help retain “its” ambiguity. Our results from Themes 3 suggest that participants enjoyed the sound as it helped them to know when to give up control while retaining the suspense of what gesture the EMS hand might show. 

Previous research showed that feedback should be given to inform the participant when the computer will take control of their body \cite{Benford_Ramchurn_Marshall_2020}. Our results support and extend this theory to the Body as a Play Material approach. We suggest that designers incorporate sound to notify players when the computer is about to take bodily control without specifying the nature of this control.

\subsubsection{Allow Players to Manipulate the Rules in Response to the Computer Control’s Limitations to Facilitate Referential Ambiguity}\label{sec7.2.2:allowmanipulations}
Huizinga \cite{Huizinga_2016} argues that game rules operate within the context of play but also suggests that players should be able to alter these rules. Our study suggested that participants had difficulty accurately calibrating the electrodes on their body to perform some gestures required for playing the games~\cite{Li2022gesplayer}. Particularly, it was those gestures that they felt uncomfortable performing. To address this, some participants changed the “air” gesture to closing the ring finger instead of the middle-finger (Fig. \ref{fig:teaser}). They changed the rules by adjusting the semantic reference of the designed gestures to something their body could perform using the EMS. This ability to manipulate rules empowered the participants to feel more in control of the gameplay.

Sutton-Smith \cite{Sutton-Smith_2001} refers to this ability to modify the rules to match the participant's abilities as "referential ambiguity". We embrace this concept of referential ambiguity and extend it to the Body as a Play Material approach. Specifically, we suggest that designers consider allowing players to manipulate the rules in response to the computer control’s limitations to facilitate referential ambiguity.

\subsubsection{Balance the Computer’s Ability to Perform Fine- and Gross-motor Movements to Amplify “Its” Ambiguity}\label{sec7.2.3:balancemovements}
The balance of gestures is crucial in digital game design for fair and engaging gameplay \cite{Jaffe_2013,Salen_Zimmerman_2004}. Participants suggested that they tried to feel the electricity in their muscles for anticipating the computer-controlled actions before the EMS hand displayed the result. Specifically, participants could more easily anticipate the one gross movement gesture of the EMS hand in Elements and Numbers compared to the two fine gestures. This ability to anticipate the gross gesture easily could be due to the imbalance between the number of fine and gross movement gestures in these games.

Previous research has found that balancing bodily games can be achieved by increasing the player's or the technology's ability in the environment \cite{Mueller_Vetere_Gibbs_Edge_Agamanolis_Sheridan_Heer_2012,Altimira_Mueller_Lee_Clarke_Billinghurst_2014,Li_Huang_Patibanda_Mueller_2023}. Our study supports this theory and extends it to the Body as a Play Material approach by suggesting that designers could consider balancing the computer’s ability to perform fine- and gross-motor movements to amplify “its” ambiguity.

\section{LIMITATIONS AND FUTURE WORK}\label{sec8:limitations}
While this work offers unique insights, it is also subject to certain limitations that future research might address. One limitation is the sample size (n=12) used in the field study, which may mean that the results are generalisable to a larger population. Moreover, our study's exploratory and descriptive nature may be prone to novelty effects \cite{Poppenk_Köhler_Moscovitch_2010}. However, our study was conducted over seven days, and we showed that our participants engaged with the system for a significant time, which allowed us to gather insights beyond the system’s novelty. Future studies with a larger sample size and a longer-term engagement or a comparative study with conventional game controls could help reduce novelty effects and provide more insights into the potential of the proposed approach. Researchers could also use quantitative game experience tools such as the player experience inventory to gather further insights about autonomy and ease of control \cite{Abeele_Spiel_Nacke_Johnson_Gerling_2020}.

One significant area identified for future research is the comfort and control associated with the EMS. The use of EMS can be discomforting for some players, and the loaning of bodily control to the technology may negatively influence the gaming experience \cite{Knibbe_Alsmith_Hornbæk_2018,Lopes_Ion_Mueller_Hoffmann_Jonell_Baudisch_2015}. Furthermore, the calibration process for EMS might not be universally effective, resulting in unexpected modifications of body movements \cite{Knibbe_Alsmith_Hornbæk_2018}. Nonetheless, we learned that designing multiple games can address this constraint to a certain extent. Future research could explore using more sophisticated EMS systems that can automate the calibration process (e.g., \cite{Knibbe_Freire_Koelle_Strohmeier_2021}) or other technologies such as pneumatics or exoskeletons that do not depend on the player’s muscular structure, possibly dampening the effects of discomfort when loaning control.

We also acknowledge that our study focused primarily on single-player, competitive games. Future research can consider employing our proposed approach to cooperative games, for example, by creating teams of EMS-controlled hands across multiple players’ bodies. Some of our participants indicated such possibilities, noting their enjoyment in putting the mobile sleeve onto another person, retaining the EMS device on their arm, and then playing the games.

In conclusion, by addressing the limitations and building upon the opportunities for future work outlined in this paper, we hope to inspire game researchers and designers to continue exploring the potential of using the body as a play material. This, in turn, could lead to the development of innovative gaming experiences that foster bodily play and engage a wider range of users in unique and inclusive ways.

\section{CONCLUSION}\label{sec8:conclusion}
In conclusion, our work proposes the Body as a Play Material approach, where players use the body as input and output by loaning bodily control to technology. This approach aims to bridge the gap by unifying the physical body and the virtual world. We designed three novel games to showcase this approach and studied them through a field study, resulting in four player experience themes. By reflecting on these themes, we provided design considerations for promoting variety and using the body as an ambiguous play material when designing future games. Ultimately, our approach brings a new perspective to creating games that aim to unify the physical body and the virtual world.

\begin{acks}
Rakesh Patibanda, Aryan Saini, Elise van den Hoven, and Florian ‘Floyd’ Mueller thank the Australian Research Council (Discovery Project Grant - DP190102068). We would also like to thank our participants who volunteered for the study. Florian ‘Floyd’ Mueller also thanks the Australian Research Council for DP200102612 and LP210200656. Their willingness to engage with our research and contribute their experiences and perspectives significantly enriched our findings. Additionally, we thank those participants who kindly agreed to be featured in the photographic material included in this publication. Explicit consent was obtained to use their images.
\end{acks}

\bibliographystyle{ACM-Reference-Format}
\bibliography{auto-paizo}

\newpage
\appendix

\section{Game Challenges}

\begin{table}[h]
  \caption{Compulsory and optional challenges were added to the software application as part of the Auto-Paízo games study.}
  \label{tab:gamechallenges}
  \begin{tabular}{>{\centering\arraybackslash}p{0.4\linewidth}>{\centering\arraybackslash}p{0.5\linewidth}}
    \toprule
    \textbf{Mandatory Challenges} & \textbf{Optional Challenges} \\
    \toprule
    Calibrate for Numbers game & Play these games in a public place \\
    \midrule
    Calibrate for Elements game & Play standing outside your house where people can see you playing \\
    \midrule
    Calibrate for Slap-Me-If-You-Can & Take it to the restaurant and see if other people are interested in playing the games \\
    \midrule
    Play while you’re on a Zoom call and other people are watching you play & Design a game by integrating these games with another game or create a new game \\
    \midrule
    Play by asking your partner, friend, or family to wear the arm guard with the phone and then let them play with your computer hand & Use these EMS games to settle a bet \\
    \midrule
    Play using both your hands (Switching calibration) & \\
    \midrule
    Integrate these games into a daily life activity (e.g., playing board games or as a decision-making tool) & \\
    \bottomrule
  \end{tabular}
\end{table}

\end{document}